\documentclass{article}

\usepackage{arxiv}

\usepackage[utf8]{inputenc} 
\usepackage[T1]{fontenc}    
\usepackage{hyperref}       
\usepackage{url}            
\usepackage{booktabs}       
\usepackage{amsfonts}       
\usepackage{nicefrac}       
\usepackage{microtype}      
\usepackage{lipsum}
\usepackage{graphicx}
\graphicspath{ {./images/} }

\title{Advanced Multi-Mode Phase Retrieval for Dispersion Scan}

\author{
 Alex M. Wilhelm \\
  Department of Physics\\
  Colorado School of Mines\\
  Golden, CO 80401 \\
  \texttt{amwilhelm@mines.edu} \\
   \And
 David D. Schmidt \\
  Department of Physics\\
  Colorado School of Mines\\
  Golden, CO 80401 \\
  \texttt{daschmid@mines.edu} \\
  \And
 Daniel E. Adams \\
  Department of Physics\\
  Colorado School of Mines\\
  Golden, CO 80401 \\
  \texttt{daadams@mines.edu} \\
    \And
 Chalres G. Durfee \\
  Department of Physics\\
  Colorado School of Mines\\
  Golden, CO 80401 \\
  \texttt{cdurfee@mines.edu} \\
}

\begin{document}
\maketitle
\begin{abstract}
We present a phase retrieval algorithm for dispersion scan (d-scan), inspired by ptychography, which is capable of characterizing multiple mutually-incoherent ultrafast pulses (or modes) in a pulse train simultaneously from a single d-scan trace. In addition, a form of Newton's method is employed as a solution to the square root problem commonly encountered in second harmonic pulse measurement techniques. Simulated and experimental phase retrievals of both single-mode and multi-mode d-scan traces are shown to demonstrate the accuracy and robustness of the algorithm.
\end{abstract}

\section{Introduction}

Over the last 50 years, ultrafast lasers have become ubiquitous tools in many fields of science, industry, and medicine \cite{Chung2009,Malinauskas2016,Dawson1979}. Their extremely short pulse durations and high peak powers make them ideal for diverse applications such as eye surgery \cite{Chung2009}, laser machining \cite{Malinauskas2016},  and particle acceleration \cite{Dawson1979,Sprangle1988,Wilhelm2019}. However, the extreme time scales of ultrafast pulses, typically less than 100 femtoseconds,  have historically made measuring their amplitude and phase profiles  difficult. To address this problem, numerous techniques have been developed to measure both the intensity and phase of ultrafast pulses. Commonly implemented techniques include, but are not limited to, frequency resolved optical gating (FROG) \cite{Trebino1993}, spectral phase interferometry for direct electric-field reconstruction (SPIDER) \cite{Iaconis1998}, multiphoton intrapulse interference phase scan (MIIPS) \cite{Lozovoy2004}, spectral phase and amplitude retrieval and compensation (SPARC) \cite{AllendeMotz:19}, and dispersion scan (d-scan) \cite{Miranda2012-1,Miranda2012-2}. 

Recently, d-scan has emerged as an experimentally simple alternative to other pulse measurement techniques. In d-scan, the pulse compressor itself is used to iteratively vary the dispersion of the pulse, and at each dispersion setting a second harmonic (SH) spectrum is recorded to generate a 2D trace. This makes d-scan an extremely simple technique to implement experimentally as the only components needed in addition to the dispersive element are a focusing mirror, a nonlinear medium, and a spectrometer. D-scan has been demonstrated in a number of experimental configurations \cite{Miranda2012-1,Miranda2012-2,Loriot2013,Fabris2015,Miranda2017,Wilhelm:20-1,Wilhelm:20-2,2021arXiv210312601G} and recently Escoto \textit{et al.} compared the performance of various phase retrieval algorithms for d-scan \cite{Escoto:18}.

A useful way of categorizing phase retrieval algorithms is to place them into one of two classes, Class I and Class II. Class I algorithms use a forward propagation model (FPM) to simulate the propagation of a guess pulse through the measurement system. The simulated guess trace is then compared to the measured trace, and various methods are used to adjust the guess pulse to minimize the error between the simulated and measured traces. Class II algorithms follow the same FPM, but apply a data constraint (projection). The FPM-constrained guess pulse is then improved by applying a second constraint (projection) after a backwards propagation model (BPM) is used. Early successful methods for retrieving the pulse from a d-scan trace followed the first class of algorithms and used the Nelder-Mead or machine-learning methods to iteratively improve the guess pulse \cite{Miranda2012-1,Miranda2012-2}. A concrete example of a Class II type algorithm is the principal components generalized projections (PCGP) algorithm for second harmonic generation (SHG) FROG with two identical pulses (i.e. \textit{not} TREEFROG \cite{10.1117/12.175862}). In this case, the algorithm uses a FPM, data constraint (measured FROG trace), BPM, and then applies a mathematical constraint (mathematical model for the SHG process) to iteratively update the guess pulse \cite{Kane2008}. More recently, ptychographic algorithms have been also been developed to address the phase retrieval problem for SHG FROG with two identical pulses \cite{Sidorenko2016}.

One of the difficulties shared by many second harmonic generation (SHG) phase retrieval algorithms that is that a square root operation must be performed to recover the fundamental field. This can lead to an ambiguity in the sign of the fundamental field as either the positive or negative root must be chosen. When the wrong root is chosen, retrieval algorithms can stagnate at a solution for the field that is incorrect. One of the advances we present in this article is a method to resolve this square-root ambiguity \cite{Jafari:19} in the context of Class II algorithms wherein no additional algorithmic steps are needed \cite{Wilhelm:20-1,Wilhelm:20-2}. Our approach is easily extended to higher-order nonlinearities, such as third-harmonic generation (THG). Following a presentation of our preliminary results \cite{Wilhelm:20-1,Wilhelm:20-2}, the THG extension of the algorithm has recently been experimentally demonstrated \cite{2021arXiv210312601G}.

Another issue inherent in pulse measurement is the potential presence of multiple mutually-incoherent pulses (modes) in a stable pulse train which alters the measured trace. We refer to such traces as multi-modal. Standard phase retrieval algorithms often fail to retrieve any of the pulses in multi-modal traces due to the assumption that the trace is generated from a single pulse. The challenge of pulse retrieval when the input pulse train is unstable has generated considerable interest \cite{VANSTRYLAND197993,Ratner:12,Escoto2019}. One objective is to ensure that a misleading retrieval is not obtained as a result of the instability: for example, algorithms have been developed for FROG that attempt to address this issue in the case where one of the retrieved modes is the coherent artifact \cite{Escoto2019}. Another objective is to quantify the level of instability by detecting the presence of pulses that add incoherently in the trace. A recent example in the context of d-scan is Escoto \textit{et al}., where several approaches were taken with a Class I differential evolution algorithm to arrive at a measure of pulse train fidelity.  \cite{Escoto2019-2}.

\begin{figure}[h]
    \centering
    \includegraphics[width=\textwidth]{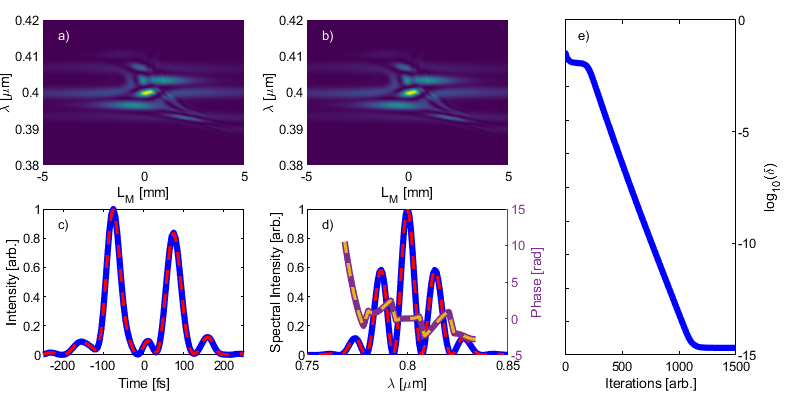}
    \caption{Simulated coherent double pulse retrieved to near machine precision ($\delta = 2\times 10^{-15}$). The delay between the pulses is $\Delta t = 150$ fs. a) Simulated d-scan trace b) Retrieved d-scan trace c) Simulated (blue) and retrieved (red-dashed) intensity profiles d) Simulated (blue) and retrieved (red-dashed) spectrum of the pulses as well as the simulated (purple) and retrieved (yellow-dashed) spectral phases e) Retrieval error $\delta$ calculated as the normalized root mean square error between the simulated and retrieved d-scan traces. }
    \label{fig:DP}
\end{figure}

Here, we present a Class II phase retrieval algorithm, inspired by ptychography \cite{Thibault2009,Maiden2009}, as a solution to both the square root problem and the multi-modal problem presented above. To this end, Fig. \ref{fig:DP} shows an example single-mode simulation of the algorithm retrieving the spectral phase and amplitude of a simulated coherent double pulse to near machine precision. In cases where the pulses are delayed well beyond the spectral resolution of the measurement device, the pulses are spectrally mutually-incoherent and the multi-mode algorithm discussed in Sec. \ref{Phase Retrieval Algorithm} can be employed. The algorithm applies Newton's method \cite{Wilhelm:20-1,Wilhelm:20-2,2021arXiv210312601G} to unambiguously determine the correct sign for the field amplitude. It utilizes the multi-modal ptychographic imaging phase retrieval concepts found in \cite{Thibault2013} to solve for a finite number of unique modes found in the pulse train. The performance of the algorithm is further explored both numerically and experimentally using SHG grating-compressor d-scan in the following sections. It is important to note that the advances of our algorithm can be applied not only to other approaches to varying the dispersion (e.g. chirp scan) but also to other pulse measurement techniques that ptychographic algorithms can be applied to (e.g. ptychographic FROG). The algorithm is robust to both noise and shot-to-shot pulse train instabilities as demonstrated in the supplemental information.

\section{Phase Retrieval Algorithm} \label{Phase Retrieval Algorithm}

The field of an ultrafast pulse can generally be represented in the spectral domain by 

\begin{equation}
    \label{eq:spectralField}
        \tilde{E}(\omega) = \tilde{E}_0(\omega)\exp(i\phi(\omega)).
\end{equation}
In d-scan, we aim to characterize both the amplitude $\tilde{E}_0(\omega)$ and spectral phase $\phi(\omega)$ of the pulse by iteratively applying known increments of spectral phase to the pulse and measuring a SH spectrum for each applied spectral phase. In the work presented here, the applied spectral phase comes from adjusting the face-to-face separation $L$ of a two grating pulse compressor. The grating compressor phase is given by \cite{Durfee1998}

\begin{equation}
    \label{eq:gPhase}
    \Phi_g(\omega,L) = \frac{\omega L}{c}\sqrt{1-\Big(\frac{2\pi c}{\omega d} - \sin{(\theta_{i})}\Big)^2}.
\end{equation}
Here, $d$ is the grating groove spacing, $\theta_{i}$ is the incident angle on the first grating, and $c$ is the speed of light in vacuum. By iteratively altering the grating separation and recording an SH spectrum at each separation, a two-dimensional trace is formed with the second harmonic wavelength on the vertical axis and the altered dispersion (i.e. grating separation) on the horizontal axis. That is to say, each column in the d-scan trace represents a SH spectrum that was measured with the pulse compressor set to a different grating separation. More generally, a measured d-scan trace can be the result of the superposition of SH spectra from multiple pulses that are mutually-incoherent. A single SH spectrum measured by the detector, with a grating separation of $L$, in such a trace can be modeled by

\begin{equation}
    \label{eq:SHGMMInt}
    \tilde{\psi}(\omega,L) = R(\omega)\Bigg[\sum\limits_{m=1}^M\Big|\int\Big(\int\tilde{E}_m(\Omega)\exp{(i \Phi_g(\Omega,L))}\exp{(i\Omega t)}d\Omega\Big)^2\exp{(-i\omega t)}dt\Big|^2\Bigg]
\end{equation}
where $R(\omega)$ is a frequency-dependant weighting factor that includes the spectral response of the doubling crystal, filters and spectrometer. Here, $M$ is the total number of incoherent modes in the trace. This equation describes the propagation of each mode through the compressor, the SHG process, propagation of the SHG field to the spectrometer, the superposition of all the modes, and measurement of the total incoherent SHG spectrum by the spectrometer. Using this model for the propagation of each mode through the system, we can develop a phase retrieval algorithm to invert the d-scan trace to retrieve the spectral phase and amplitude of each mode in the trace simultaneously.

Our algorithm is inspired by ptychography, a coherent diffractive imaging (CDI) technique. In ptychography, a probe beam is iteratively scanned across an object in a series of overlapping spatial positions. At each position, a diffraction pattern is recorded on a detector. An iterative phase retrieval algorithm can then be used to reconstruct the spatial amplitude and phase of both the object and the probe beam \cite{Thibault2009,Maiden2009}. The basic idea of our algorithm is inspired by previous ptychographic algorithms applied to both FROG and d-scan \cite{Miranda2017,Sidorenko2016}, which were adapted from the extended Ptychographic Iterative Engine (ePIE) to retrieve a single coherent pulse \cite{Maiden2009}. Our algorithm uses a simple model of the pulse propagation from the compressor to the spectrometer and back, along with a data constraint, to extract the spectral phase and amplitude of the pulse. 

\begin{figure}
    \centering
    \includegraphics[width=0.8\textwidth]{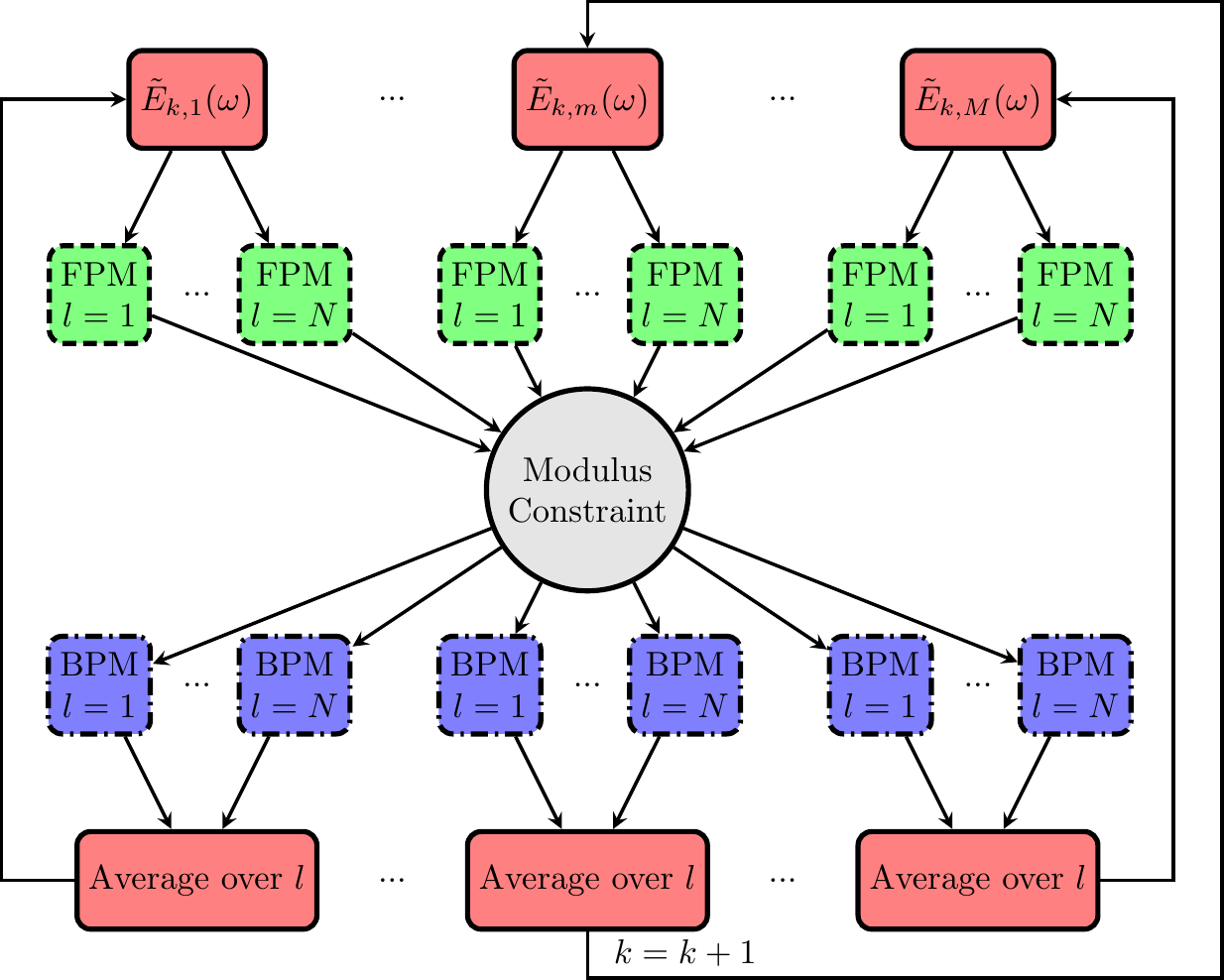}

    \caption{Flowchart of the multi-modal phase retrieval algorithm. The red solid-bordered boxes represent the beginning and end of each iteration of the algorithm. The green dash-bordered boxes represent the FPM for each of column of the trace. The grey circle represents the multi-modal modulus constraint. The blue dash-dot-bordered boxes represent the BPM for each column in the trace.}
    \label{fig:flowchart}
\end{figure}

The algorithm can be broken up into four primary components: the FPM, the modulus constraint (MC), the BPM, and a weighted average over each constrained guess. The algorithm begins by generating a guess field $\tilde{E}_{k,m}(\omega)$ for each of the $M$ modes present in the trace. In this notation, $k$ refers to the current iteration of the main algorithm loop and $m$ refers to one of the $M$ modes present in the trace. A diagram of the algorithm is shown in Fig. \ref{fig:flowchart}.

The FPM generates the calculated d-scan trace from the current guess by simulating the propagation of the pulse from the output of the pulse compressor, through the doubling crystal and to the spectrometer. This is accomplished by applying the phase from the pulse compressor at each of the $N$ grating separations to $\tilde{E}_{k,m}(\omega)$ simultaneously:

\begin{equation}
    \label{eq:Xiw}
    \tilde{X}_{k,l,m}(\omega) = \tilde{E}_{k,m}(\omega)\exp{(i\Phi_g^l)}.
\end{equation}
Here, $l$ indicates which of the $N$ grating phases is being applied to the guess field. This new set of fields are then propagated through the doubling crystal, under the assumption that the doubling process is lossless and instantaneous, to produce a set SHG fields in the time domain:

\begin{equation}
    \label{eq:Psiit}
    \psi_{k,l,m}(t) = (\mathcal{F}^{-1}\{\tilde{X}_{k,l,m}(\omega)\})^2.
\end{equation}
The SHG fields are then propagated through the spectrometer:

\begin{equation}
    \label{eq:Psiiw}
    \tilde{\psi}_{k,l,m}(\omega) = \mathcal{F}\{\psi_{k,l,m}(t)\}.
\end{equation}
In the spectrometer plane, the MC replaces the amplitude of each SHG field $\tilde{\psi}_{k,l,m}(\omega)$ with the amplitude of the measured SHG spectrum, $\tilde{T}_l$ at the $l^{th}$ grating separation while the phase of the guess is preserved:

\begin{equation}
    \label{eq:PsiiwPrime}
    \tilde{\psi}_{k,l,m}'(\omega) = \left[\frac{\tilde{T}_l}{\sum\limits_{m=1}^M|\tilde{\psi}_{k,l,m}(\omega)|^2}\right]^{1/2}\tilde{\psi}_{k,l,m}(\omega).
\end{equation} 
This modulus constraint forces the updated SHG fields to be constrained by the amplitude of the measured SHG spectrum while enforcing the notion that the trace is formed by the superposition of multiple incoherent modes. It can easily be seen that in the presence of a single mode in the trace, that this modulus constraint reduces to ones shown previously in the literature \cite{Miranda2017,Sidorenko2016}.

Now, in the BPM, the constrained SHG fields are then propagated back to the crystal plane:

\begin{equation}
    \label{eq:PsiitPrime}
    \psi_{k,l,m}'(t) =\mathcal{F}^{-1}\{\tilde{\psi}_{k,l,m}'(\omega)\}.
\end{equation}
At the crystal plane, the the SHG field is effectively propagated back through the crystal to obtain the fundamental fields just after the compressor:

\begin{equation}
    \label{eq:XitPrimeMax}
    X_{k,l,m}'(t) = \frac{1}{2}\Bigg[2 X_{k,l,m}(t)+\frac{X_{k,l,m}^*(t)}{|X_{k,l,m}(t)|^2_{max}}(\psi_{k,l,m}'(t)-\psi_{k,l,m}(t))\Bigg].
\end{equation}
The $max$ operation is taken to improve the stability of the algorithm by avoiding singularities in the field. Equation \ref{eq:XitPrimeMax} will be discussed in more detail in Section \ref{sec:Newton}. The new fundamental fields are then propagated back through the compressor to obtain $N$ new guesses for the field for each mode:

\begin{equation}
    \label{eq:EiwPrime}
    \tilde{E}_{k,l,m}'(\omega) = \mathcal{F}\{X_{k,l,m}'(t)\}e^{-i\Phi_g^l}.
\end{equation}

The final step of the algorithm is to generate a single initial guess for each mode for the next iteration of the loop. This must be done as each $\tilde{E}_{k,l,m}'(\omega)$ only contains information from a single column of the trace, and ultimately a single unique solution must be found for the field for each mode. To generate a single new guess for $k+1$ iteration, we apply a weighted average to $\tilde{E}_{k,l,m}'(\omega)$ along $l$:

\begin{equation}
    \label{eq:weightedAverage}
    \tilde{E}_{k+1,m}(\omega) = \frac{\sum\limits_{l=1}^N \tilde{E}_{k,l,m}'(\omega)}{\sum\limits_{l=1}^N \big|e^{i \Phi_g^l}\big|^2} = \frac{1}{N}\sum\limits_{l=1}^N \tilde{E}_{k,l,m}'(\omega).
\end{equation}
The averaging step mixes information from all parts of the trace to inform the next iteration of the loop. More accurately, it tends to help enforce the notion that the sub-trace from each mode is generated from a set of identical pulses that only differ by some known dispersion (i.e. that the underlying pulse is not changing from column to column). This leads to vastly improved convergence stability in noisy traces where the edges of the trace might have low signal-to-noise (SNR) ratios, in traces where spectral interference is present due to multiple coherent pulses within a mode, or when the pulse train itself is inherently unstable.

The new guesses for the fields $\tilde{E}_{k+1,m}(\omega)$ are then fed into the FPM as the start of the next iteration of the loop. The loop will repeat until a desired threshold error (discussed later) is met. 

\section{Newton's Square Root Method}
\label{sec:Newton}

In the previous section, the purpose of Eq. \ref{eq:XitPrimeMax} is to determine the fundamental field from the SHG field. A form of this equation is often directly transferred from similar ptychography algorithms. In diffractive imaging, the probe and object are unique with no ambiguities. However, in the case of the SHG model used here, this equation amounts to taking the square root of the SHG field in the time domain. This square root leads to an issue in the phase retrieval problem due to an ambiguity in the overall sign of the root:

\begin{equation}
    \label{eq:sqrt}
    \sqrt{z} = \pm\sqrt{r}e^{i\theta/2}.
\end{equation}
One of the two possible roots must be chosen for the fundamental field, but without the ability to directly measure the sign of the field, it is for the algorithm to determine which root is correct. Standard phase retrieval algorithms typically solve this problem by simply choosing the principle root. In many cases this is an adequate solution, but it is not necessarily the correct solution. In simulation, we have found that choosing the principal root can lead to stagnation of phase retrieval algorithms in local minima when the true fundamental field is not the principal root of the SHG field. As real fields can be either root, it is prudent to use a phase retrieval algorithm that will naturally choose the correct root without any prior knowledge of the fundamental field and without additional measurements. We assert that Eq. \ref{eq:XitPrimeMax} solves this square root problem by not strictly enforcing either of the two roots. Instead, the modulus constraint, in combination with the averaging appearing in Eq. \ref{eq:weightedAverage}, essentially chooses the correct root, and Eq. \ref{eq:XitPrimeMax} simply allows that choice to be preserved.

To see this, we can compare Eq. \ref{eq:XitPrimeMax} to Newton's method for calculating the $n^{th}$ root of a number \cite{NewtonFluxions}. In Newton's method, given a number $A$ and an initial guess for its $n^{th}$ root $x_i$, the next guess is given by:

\begin{equation}
    \label{eq:nRoot}
    x_{i+1} = x_i-\frac{1}{n}\Bigg(x_i-\frac{A}{x_i^{n-1}}\Bigg).
\end{equation}
Using this method, either the positive or negative roots can be found by starting with a positive or negative guess, respectively. For $n=2$ in our notation, this is equivalent to:

\begin{equation}
    \label{eq:sqrtSimple}
    X_{k,l,m}'(t) = \frac{1}{2}\Bigg[X_{k,l,m}(t)+\frac{\psi_{k,l,m}'(t)}{X_{k,l,m}(t)} \Bigg].
\end{equation}
By adding 0 to the right side of this equation in the form of
\begin{equation}
    \label{eq:zero}
    0 = X_{k,l,m}(t) - \frac{X^*_{k,l,m}(t) \psi_{k,l,m}(t)}{X^*_{k,l,m}(t) X_{k,l,m}(t)},
\end{equation}
we arrive back at Eq. \ref{eq:XitPrimeMax} (with the exception of the \textit{max} operation). It follows then that the modulus constraint and the averaging appearing in Eq. \ref{eq:weightedAverage} choose the correct root for the field, and Eq. \ref{eq:XitPrimeMax} simply preserves that choice. Similar equations have been used before in FROG retrieval algorithms to update the fundamental field guess using the SHG field, but they make use of a weighting factor $\alpha \in [0,1]$ to control the strength of the field update \cite{Sidorenko2016,Miranda2017}. After comparing to Newton's method, we find that the optimal choice for $\alpha$ in these methods is 1/2 as it amounts to exactly taking the square root and also seems to provide the best balance between speed and stability for these algorithms. To our knowledge, this is the first time the connection between the ptychographic field update and Newton's square root method has been identified for SHG phase retrieval techniques.

Additionally, this analysis can be extended to other perturbative nonlinearities such as third harmonic generation (THG). Applying the same methodology as above to THG techniques gives

\begin{equation}
    \label{eq:THGUpdate}
    X_k(t) = X_k(t) + \frac{1}{3}\frac{(X_k(t)^*)^2}{|X_k(t)|^4}(\xi_k'(t)-\xi_k(t))
\end{equation}
as the method for taking the cube root in a THG pulse retrieval algorithm. Here, $\xi$ is the THG field and $\xi'$ is the THG field with the modulus constraint applied.

\section{Numerical Results}
\subsection{Single Mode Simulations}
We now demonstrate the performance of the retrieval algorithm on simulated single-mode d-scan traces in order to evaluate the algorithm's performance in the typical scenario where a single coherent mode is present in the trace. All traces were simulated on a grid containing 512 spectral points and 500 grating positions evenly spaced over 20 mm. Pulses in the simulations were generated from Gaussian spectra centered at $\lambda_0 = 800$ nm with a full width at half maximum (FWHM) bandwidth of $\Delta\lambda = 31.4$ nm which can support pulses as short as $\Delta\tau = 30$ fs. However, we note that we observed that the algorithm retrieved any physical pulse we simulated. The simulation and retrieval grating phases were each modeled using 1400 lines/mm gratings with Littrow incident angle ($\theta_i = 34.05^{\circ}$). For noiseless traces, simulations were run until the NRMSE 

\begin{equation}
    \label{eq:RMS}
    \delta = \sqrt{\frac{\sum\limits_{\omega,l}(S_{\omega,l}-R_{\omega,l})^2}{N P}}
\end{equation}
between the simulated (or measured), $S$, and retrieved, $R$, traces approached machine precision ($\delta \sim 10^{-16}$). Here, $P$ is the total number of spectral points. For traces with added noise, simulations were run until $\delta$ stagnated at a minimum value.

\begin{figure}[h]
    \centering
    \includegraphics[width=\textwidth]{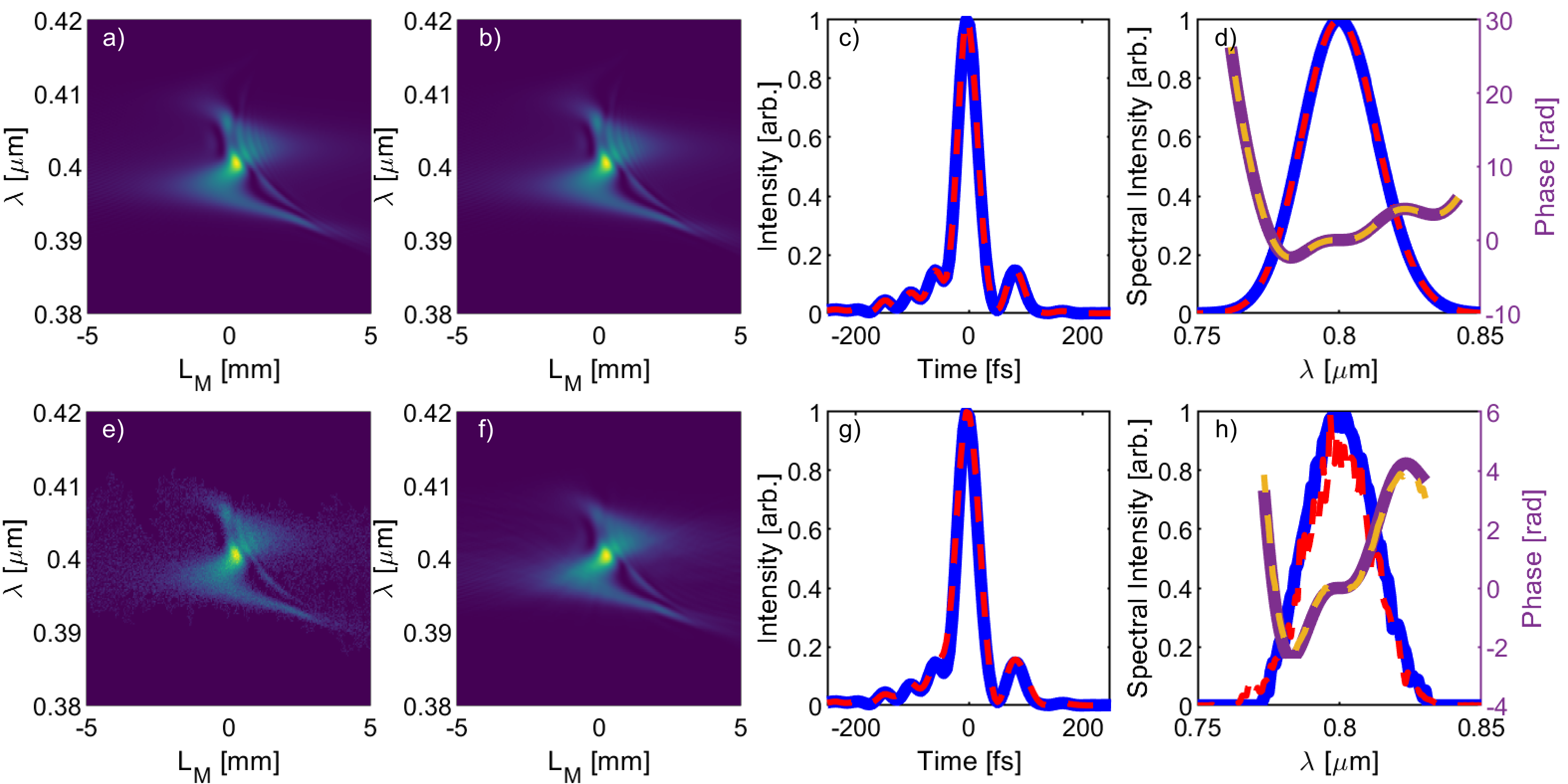}
    \caption{Simulated pulse retrievals of two different single-mode d-scan traces. The leftmost column shows the simulated d-scan traces. The middle-left column shows the retrieved traces. The middle-right column shows the simulated (blue) and retrieved (red-dashed) intensity profiles of the pulses. The right-most column shows the simulated (blue) and retrieved (red-dashed) spectra as well as the simulated (purple) and retrieved (yellow-dashed) spectral phases.}
    \label{fig:SingleMode}
\end{figure}

Figure \ref{fig:SingleMode} shows the results of two retrievals on highly-structured, simulated single-mode traces. The simulated spectral phase content of the pulses in these simulations consist of third and fourth order phase as well as sinusoidal phase described by

\begin{equation}
    \label{eq:SimPhase}
    \phi_S(\omega) = \phi_{3,S}(\omega-\omega_0)^3 + \phi_{4,S}(\omega-\omega_0)^4+\sin(\alpha(\omega-\omega_0)).
\end{equation}
Here, $\phi_{3,S} = 3.4\times10^4\textrm{ fs}^3$, $\phi_{4,S} = 1\times10^6\textrm{ fs}^4$, and $\alpha = 75$ fs. The retrieval in Fig. \ref{fig:SingleMode}(a) - \ref{fig:SingleMode}(d) shows the result of a retrieval on a trace generated from a single pulse with the spectral phase in Eq. \ref{eq:SimPhase}. In this simulation the trace, temporal intensity, spectrum, and spectral phase were all retrieved to machine precision.

Figure \ref{fig:SingleMode}(e)-\ref{fig:SingleMode}(h) shows the pulse retrieval on a trace with additive white Gaussian noise (AWGN) added to the trace. The AWGN for this simulation had a signal-to-noise ratio of 25. We will define the signal to noise ratio (SNR) as the peak intensity of the trace divided by the standard deviation of the background noise:

\begin{equation}
    \label{eq:SNR}
    SNR = \frac{S_{max}}{\sigma_{bkrnd}}.
\end{equation}
The retrieved trace agrees well with the simulated trace, and the pulse is retrieved extremely accurately. While the spectral phase is very accurately retrieved across the entire bandwidth, some small distortions of the retrieved spectrum can be seen. This simulation is an extreme example of how robust the algorithm is to noise in an experiment. While the SNR in this simulation was 25, it is generally higher than 500 in typical measurements on our laser system. The supplemental information further explores how imperfections in the d-scan trace and pulse train instabilities can effect the pulse retrieved by the algorithm.

It should be noted that in order to achieve accurate retrievals in the presence of noise, the raw data requires background subtraction or noise suppression before running the retrieval algorithm. The method of background subtraction we have chosen to employ uses a median threshold on the columns of the trace. The threshold is chosen as the standard deviation of the background noise and all values below the threshold multiplied by the median of the column are set to zero.

Figure \ref{fig:DP} shows a retrieval of a trace generated from two coherent pulses with the above spectral phase but with an additional time delay of $\Delta t = 150$ fs between the two pulses. Just as with the single pulse noiseless trace, temporal intensity, spectrum, and spectral phase were all retrieved to approximately machine precision. 

\subsection{Multi-mode Simulations}

\begin{figure}[h]
    \centering
    \includegraphics[width=\textwidth]{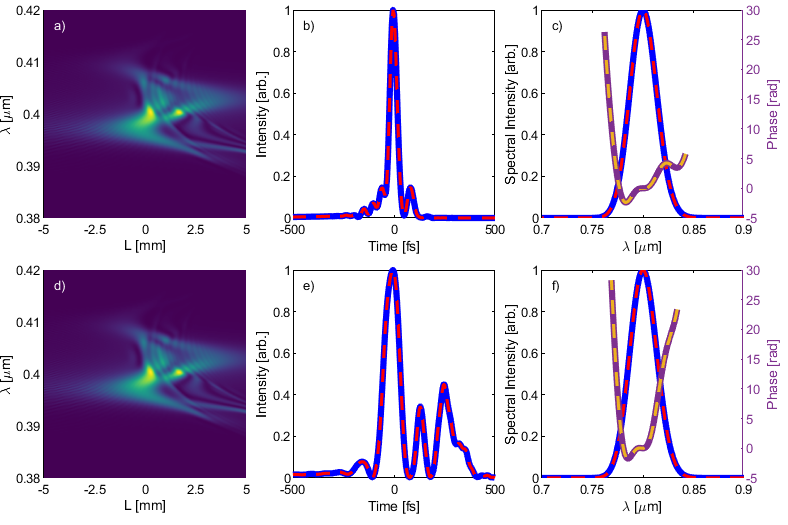}
    \caption{Simulated pulse retrievals on a multi-mode trace who's two pulses differ in dispersion by the dispersion of 75mm of BK7 glass. a) Simulated trace. b) Simulated (blue) and retrieved (red-dashed) pulse for the first mode. c) Simulated (blue) and retrieved (red-dashed) spectrum as well as simulated (purple) and retrieved (yellow-dashed) spectral phase for the first mode. d) Retrieved trace. e) Simulated (blue) and retrieved (red-dashed) pulse for the second mode. f) Simulated (blue) and retrieved (red-dashed) spectrum as well as simulated (purple) and retrieved (yellow-dashed) spectral phase for the second mode. }
    \label{fig:MMSim}
\end{figure}

In this section we numerically demonstrate the ability of the algorithm to retrieve the spectral phase and amplitude of two mutually-incoherent pulses simultaneously from a single d-scan trace. To do so, two mutually-incoherent pulses, or modes, were simulated on the same trace. The first mode was identical to the pulse in the single-mode simulations above. The second mode was the same as the first mode but with additional 75 mm of BK7 glass phase added to the pulse. This thickness of glass was chosen in order to simulate situations where the two modes are individually coherent, but the spectral interference fringes cannot be resolved by the spectrometer as the delay between the pulses is too long. The two modes were then added incoherently to generate the simulated trace. That is to say that the SHG signal in each column of the trace, $\tilde{I}(\omega)$, is the sum of the individual signals from the two modes:

\begin{equation}
    \label{eq:SHGMM}
    \tilde{I}(\omega) = \sum_{m=1}^2|\tilde{\psi}_m(\omega)|^2.
\end{equation} 
The results of running the algorithm on this simulated multi-modal trace are show in Fig. \ref{fig:MMSim}. In the simulated trace, the sub-trace from each mode can be seen stacked on top of one another. The pulse profiles, spectral phase, and spectrum were all retrieved to machine precision for both modes.

\section{Experimental Results}

\subsection{Experimental Setup}

Lastly, we experimentally verify the performance of the retrieval algorithms using a home-built Ti:Sapphire laser system. The output of the regenerative amplifier and transmission grating compressor chain delivers 0.75 mJ pulses at 1 kHz with a $\Delta\lambda = 30.2$nm FWHM bandwidth centered near $\lambda_0 = 803$ nm. Pulses were focused into a 50 $\mu$m KDP crystal using an f = 500 mm cylindrical mirror. Spectra were collected using a fiber-coupled Ocean Optics Flame spectrometer. A BG-39 bandpass filter was used to suppress fundamental light while collecting SHG spectra and was removed to collect reference fundamental spectra. A white light calibration of the spectral response of the spectrometer and filter was performed, and the phase matching response of the KDP crystal was accounted for to improve the accuracy of all experimental retrievals.

\subsection{Single Mode Retrieval}
\begin{figure}[h]
    \centering
    \includegraphics[width=\textwidth]{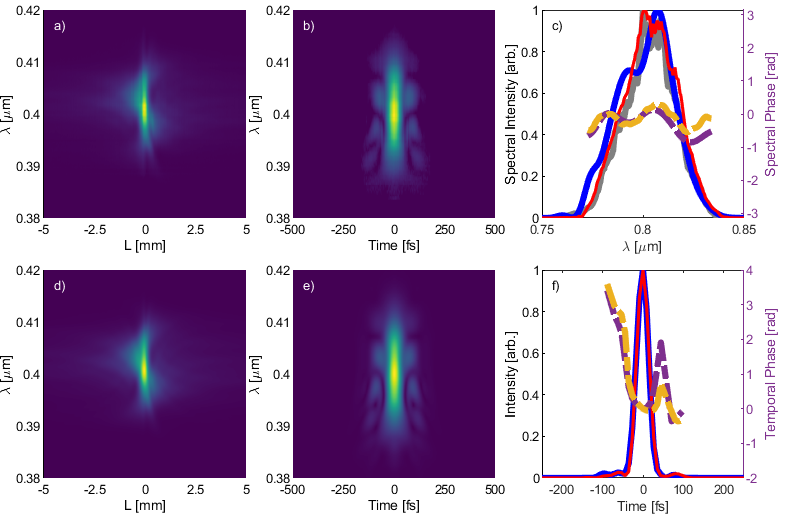}
    \caption{Experimental measurement of the pulse from our home-built laser system. a) The measured d-scan trace. b) The measured FROG trace for comparison. c) The reference measured spectrum (grey), retrieved FROG (blue) and retrieved d-scan (red) spectra, along with retrieved FROG (purple-dashed) and retrieved d-scan (yellow-dashed) spectral phase. d) Retrieved d-scan trace. e) Retrieved FROG trace. f) Retrieved FROG (blue) and d-scan (red) pulses along with retrieved FROG (purple-dashed) and d-scan (yellow-dashed) temporal phases.}
    \label{fig:FROGCompare}
\end{figure}

We demonstrate the single-mode parallel algorithm on a near transform-limited pulse from the system described in the previous section. The measured trace contains 500 grating separations over 10mm. A scanning FROG trace was also measured for comparison. The results of this experiment are shown in Fig. \ref{fig:FROGCompare}.

The retrieved pulses from both the FROG and the d-scan show excellent agreement, with a retrieved duration of $\tau = 36$ fs FWHM duration. The spectral and temporal phases from the two methods also match each other quite well. Some small differences can be seen between the retrieved spectra for the two methods, and we attribute these largely to the different optics used in the FROG and d-scan devices. However, the retrieved d-scan spectrum matches the measured reference spectrum very well. The NRMSE between the measured and retrieved d-scan traces for this retrieval was $\delta = 1.6\times10^{-3}$. 

\subsection{Multi-Mode Retrieval}

\begin{figure}[h]
    \centering
    \includegraphics[width=\textwidth]{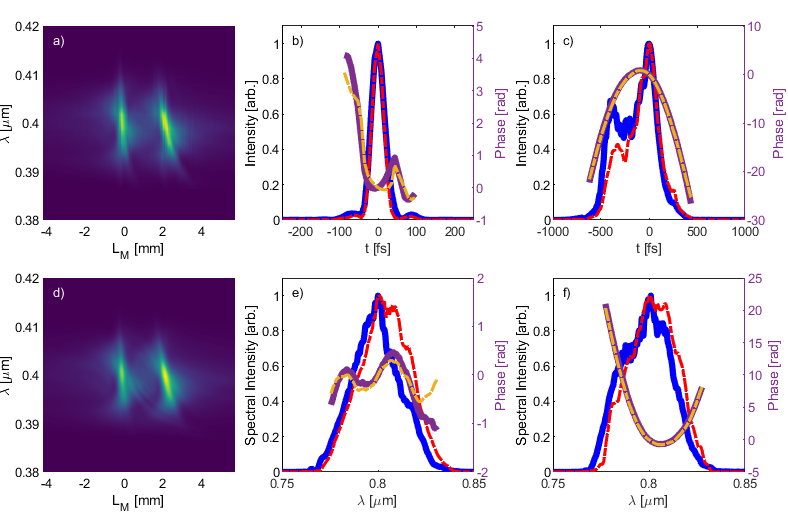}
    \caption{Retrieval results using the multi-modal algorithm from an experimentally measured multi-mode d-scan trace. a) Measured multi-mode trace. b) Retrieved pulses and temporal phases for mode 1. c) Retrieved pulses and temporal phases for mode 2. d) Retrieved multi-mode trace. e) Retrieved spectra and spectral phases for mode 1. f) Retrieved spectra and spectral phases for mode 2. The blue and purple lines correspond to the results from the multi-mode algorithm and multi-mode trace, while the red-dashed and yellow-dashed lines correspond to results from the single-mode algorithm used on the comparison single-mode traces. Mode 1 corresponds to the leftmost mode in the trace and mode 2 corresponds to the rightmost mode in the trace.}
    \label{fig:MMExperiment}
\end{figure}

Next, we demonstrate the ability of the algorithm to retrieve multiple incoherent modes simultaneously from a single measured trace experimentally. To generate a multi-mode trace, the timing of the internal Pockels in the regenerative amplifier was adjusted so that two pulses circulated simultaneously, with one of the pulses taking an extra round trip. The two pulses were separated in time by the round trip time of the amplifier, approximately 12 ns, which is well outside the typical temporal range of FROG devices. This is a similar to the above simulated example where both pulses are technically coherent, but do not temporally overlap enough for their spectral interference to be resolved by the spectrometer. A trace, shown in Fig. \ref{fig:MMExperiment}(a), was recorded while both pulses were present in the pulse train. The extra round trip significantly increases the dispersion of the second pulse, causing its SHG signal to shift along the dispersion axis of the trace. This trace was then processed by the multi-modal algorithm to retrieve both pulses in the train simultaneously, as well as their spectra and phase. 

A single-mode trace was also measured for each of the two pulses by adjusting the Pockels cell timing so that either $x$ or $x+1$ round trips were taken in the amplifier. The zero position of the compressor was kept constant for all three traces and was chosen to be the point of best compression for the mode that took fewer round trips in the amplifier. The pulses and spectra retrieved from the single-mode traces were used as a reference for the results of the multi-mode retrieval. It should be noted that though we use the single-mode retrievals as a reference for the multi-mode retrieval, we do not expect them to be exactly the same for reasons that will be discussed below. To our knowledge no other measurement technique exists that can measure both pulses simultaneously to compare against. The results of the single-mode and multi-mode retrievals can be seen in Fig. \ref{fig:MMExperiment}(b)-\ref{fig:MMExperiment}(f).

It should be noted that the multi-mode trace originally contained 1200 grating separations over a distance of 24 mm, rather than the 500 grating separations over 10 mm that the single-mode traces contained. The trace was then cropped to 501 grating positions so that the two modes were approximately centered on the cropped trace. This removed areas near the edges of the original trace that had low SNR, and ensured that the two modes were approximately equidistant from the edges of the trace. This helps ensure that the NRMSE of the retrieval monotonically decreases (i.e. does not oscillate) and that both modes are accurately retrieved.

It can easily be seen from Fig. \ref{fig:MMExperiment}(a) and \ref{fig:MMExperiment}(d) that the retrieved multi-mode trace matches the measured one quite well by eye, with only minor differences between the two. The NRMSE for this retrieval was $\delta = 1.1\times10^{-2}$. This is approximately an order of magnitude higher than the error in the single-mode demonstration above, and we attribute this primarily to the fact that the laser becomes more unstable when operating in this multi-pulse regime. 

Figure \ref{fig:MMExperiment}(b) show the retrieved pulse and temporal phase for Mode 1 from both the multi-mode and single-mode retrievals and Fig. \ref{fig:MMExperiment}(e) shows the retrieved spectrum and spectral phase for mode 1 from both the single-mode and multi-mode retrievals. The temporal and spectral phase from the multi-mode retrieval match the corresponding ones from the single phase retrieval  quite well. Additionally, the temporal profile from each retrieval is nearly identical. There is a small discrepancy between the shape of the retrieved spectra with the multi-mode spectrum being slightly narrower and blue-shifted. We attribute this spectral discrepancy to the difference in gain dynamics in the amplifier when two pulses are present rather than just one. We expect less spectral gain shaping for each pulse when both are present in the amplifier.

Figures \ref{fig:MMExperiment}(c) and \ref{fig:MMExperiment}(f) show the temporal and spectral components of the retrieval for mode 2. Again, the temporal and spectral phases agree extremely well between the one- and two-pulse retrievals. When mode 2 and mode 1 are present in the amplifier at the same time the retrieved spectrum of mode 2 is red-shifted and broader relative to mode 1 which brings it more in line with the spectrum from the single-mode retrieval as is expected. Additionally, the temporal profiles for mode 2 are slightly different between the two retrieval methods due to the different gain dynamics in the multi-mode measurement. It is more significant for this mode as it contains more energy due to the extra round trip it takes. Lastly, it should be noted that the spectral phases are not affected by the gain dynamics because the propagation in the amplifier is fully linear and they are only affected by the material dispersion of the amplifier itself.

\begin{figure}[h]
    \centering
    \includegraphics[width=\textwidth]{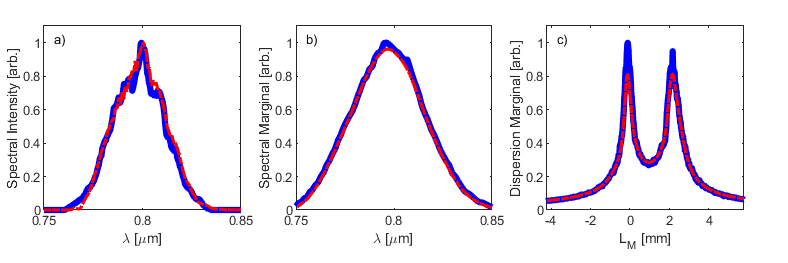}
    \caption{Further verification data from the multi-mode retrieval. a) Mutually-incoherent spectrum b) Spectral Marginal c) Dispersion Marginal. Measured quantities shown in blue lines and retrieved quantities shown in red-dashed lines.}
    \label{fig:MMExperiment2}
\end{figure}

Based on the comparisons against the single-mode retrievals, there is significant evidence that the multi-mode algorithm is accurately retrieving both pulses in the pulse train. Additional verification of the retrieval can be found by analyzing the retrieved incoherent spectrum, spectral marginal, and dispersion marginal.  A reference spectrum was measured for the multi-mode pulse train. This spectrum is the incoherent sum of the spectra from the individual pulses

\begin{equation}
    \label{eq:specSum}
    I_T(\omega) = I_1(\omega) + I_2(\omega).
\end{equation}
To compare against this measurement, we can simply add the retrieved spectra from the multi-mode retrieval together. Figure \ref{fig:MMExperiment2}(a) shows this comparison. While the individual multi-mode spectra do not perfectly match their single-mode counterparts, the retrieved incoherent spectrum matches the measured one extremely well.

The spectral marginal is the integral of the trace $\tilde{T}$ along the dispersion axis, which is parametrized by the compressor separation here

\begin{equation}
    \label{eq:specMarg}
    \Pi_{\omega}(\omega) = \int \tilde{T}(2\omega,L_M) dL_M.
\end{equation}
In Fig. \ref{fig:MMExperiment2}(b), it can be seen that the measured and retrieved spectral marginals agree extremely well. Lastly, the dispersion marginal is the integral of the trace along the spectral axis

\begin{equation}
    \label{eq:dispMarg}
    \Pi_{L_M}(\omega) = \int \tilde{T}(2\omega,L_M) d\omega.
\end{equation}
Figure \ref{fig:MMExperiment2}(c) shows a comparison between the measured and retrieved dispersion marginal. The two curves largely agree with only a small difference in the peak heights. The degree to which the incoherent spectra, and the two marginals match gives further evidence that the multi-mode algorithm accurately retrieved both pulses and their spectra from the multi-mode trace.

\section{Conclusion}

We have demonstrated, both numerically and experimentally, a ptychographic phase retrieval algorithm that is capable of solving two common problems in ultrafast pulse measurement. First, it is capable of choosing the correct root for the fundamental field with no external knowledge of the sign of the field or ancillary measurements needed. It does this through careful application of Newton's square root method to the special case of SHG. This method can also be generalized to higher order processes such as THG. Second, the algorithm can be extended to solve for multiple mutually-incoherent pulses simultaneously. This effectively allows the d-scan to measure a finite number of distinct pulses in a single measurement regardless of the time delay between the pulses. This is a powerful tool for diagnosing pulses with different dispersion characteristics in laser systems and the stability of pulse trains \cite{Escoto2019-2}. The algorithm was also shown to be robust to both inherent noise in the measurement and instabilities in the pulse train. The ability of the algorithm to retrieve simulated pulses to machine precision and to retrieve experimental pulses with low error shows that our algorithm improves on previous d-scan phase retrieval algorithms \cite{Miranda2012-1,Miranda2012-2,Miranda2017}, making it well suited for accurately retrieving the full spectral phase and amplitude of a pulse or series of pulses. It should be stated though, that special care must be taken to ensure that the dispersion of the apparatus is at least moderately well known, and that the spectral response of the apparatus (e.g. spectrometer calibration, phase matching) is correctly accounted for to achieve highly accurate retrievals. We conclude that the d-scan measurement technique, in combination with the ptychographic algorithm presented here, is a simple and effective tool for quantitatively diagnosing the pulse train from a laser system, as well as the laser system itself.

\section*{Funding}
A. Wilhelm and C. Durfee gratefully acknowledge support from the National Science Foundation under grant numbers PHY-1619518 and PHY-1903709. D. Schmidt gratefully acknowledges support under AFOSR grant FA9550-16-1-0121. D. Adams gratefully acknowledges support under AFOSR grant FA9550-18-1-0089.

\section*{Disclosures}
The authors declare no conflicts of interest.

\section*{Supplemental Information}
See the supplemental information for supporting content.

\bibliographystyle{unsrt}  


\end{document}


\title{Advanced Multi-Mode Phase Retrieval for Dispersion Scan Supplemental Information}

\author{
 Alex M. Wilhelm \\
  Department of Physics\\
  Colorado School of Mines\\
  Golden, CO 80401 \\
  \texttt{amwilhelm@mines.edu} \\
   \And
 David D. Schmidt \\
  Department of Physics\\
  Colorado School of Mines\\
  Golden, CO 80401 \\
  \texttt{daschmid@mines.edu} \\
  \And
 Daniel E. Adams \\
  Department of Physics\\
  Colorado School of Mines\\
  Golden, CO 80401 \\
  \texttt{daadams@mines.edu} \\
    \And
 Chalres G. Durfee \\
  Department of Physics\\
  Colorado School of Mines\\
  Golden, CO 80401 \\
  \texttt{cdurfee@mines.edu} \\
}

\maketitle
\begin{abstract}
In this supplementary document we analyze how various pulse train instabilities affect the retrieval of a pulse from the d-scan trace using the single-mode algorithm. Retrievals from simulated traces with simulated instabilities are shown.
\end{abstract}
\section{Pulse Train Instabilities And The Coherent Artifact}

\begin{figure}[h]
    \centering
    \includegraphics[width=\textwidth]{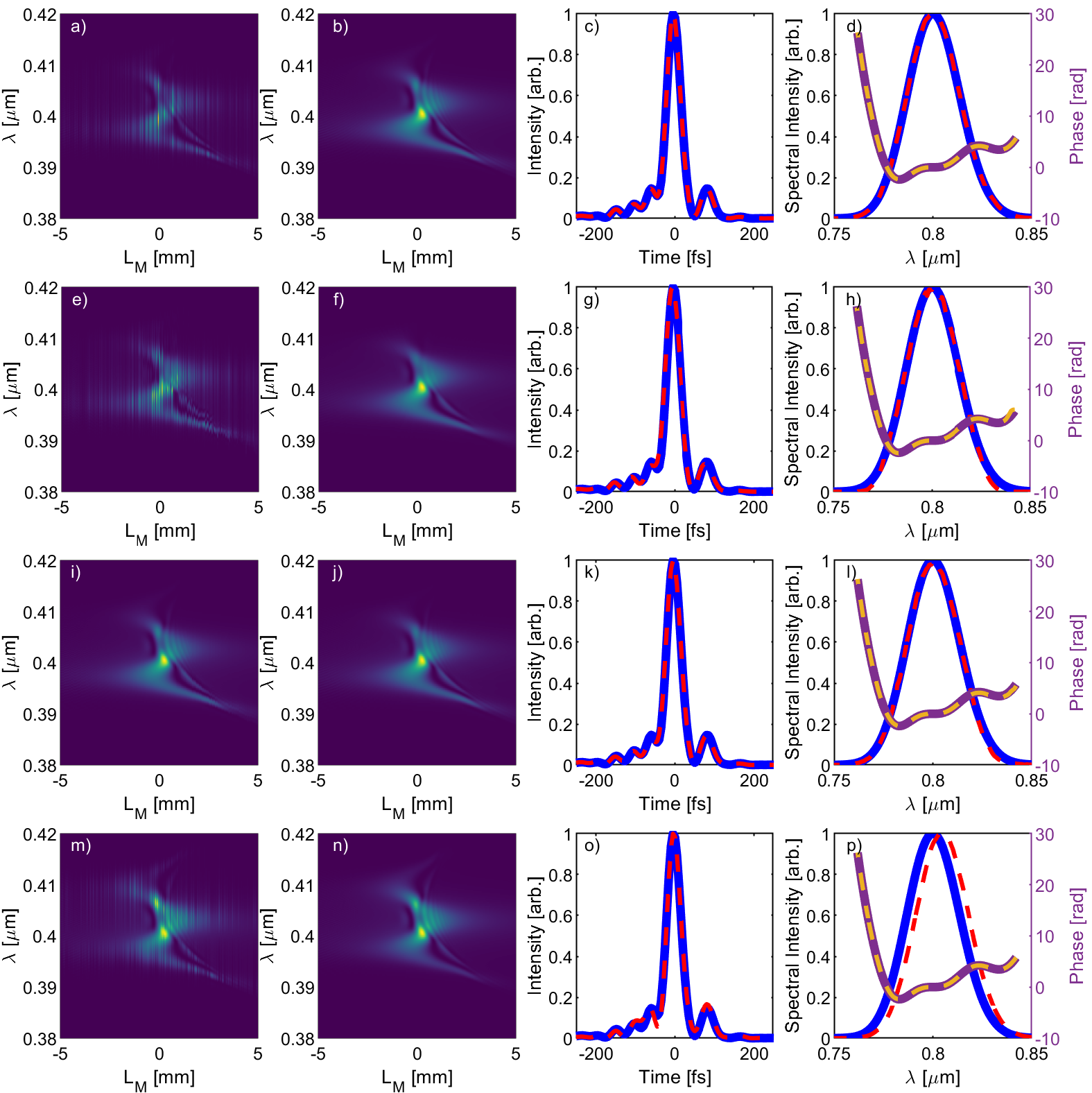}
    \caption{Simulated pulse retrievals on traces with different sources of error. The four rows correspond to the following sources of error: energy fluctuations in the pulse train, phase fluctuations in the pulse train, shot-to-shot pointing instabilities into the pulse compressor, and unstable gain pulling in the amplifier. }
    \label{fig:NoisyTraces}
\end{figure}

In most scanning dispersion scan (d-scan) measurements, thousands of pulses in a train are measured and are implicitly assumed to be identical to each other. Generally, this is a valid assumption as the pulse train is reasonably stable. However, when the pulse train is unstable, the instabilities are encoded into the trace and the retrieval algorithm faces the impossible task of trying to recover a single pulse from a trace containing information from thousands of pulses. For some measurement techniques (e.g. autocorrelation, FROG, SPIDER) retrievals can exhibit a coherent artifact: the retrieved pulse from an unstable pulse train is not representative of the average pulse in the train \cite{doi:10.1063/1.1653002,VANSTRYLAND197993,Ratner:12}.  The effect of pulse train instabilities and the coherent artifact has been extensively studied for frequency resolved optical gating (FROG) \cite{Ratner:12,Escoto2019}, and recently has been explored for other d-scan retrieval algorithms \cite{Alonso:19,Escoto2019-2,Alonso2020}. While d-scan has been shown to not exhibit a misleading coherent artifact \cite{Escoto2019-2}, it is still instructive to understand how pulse instabilities affect the d-scan trace and how our algorithm behaves in the presence of pulse train instabilities. We have identified four common sources of instabilities in laser pulse trains that we will test the algorithm on: shot-to-shot energy fluctuations, phase fluctuations, pointing instabilities within the laser, and unstable gain pulling during amplification. Figure \ref{fig:NoisyTraces} shows the results of four separate simulations, each exhibiting one of these instabilities. The leftmost column shows the simulated traces for each instability source. The middle-left column shows the retrieved traces. The middle-right column shows the ideal simulated pulse (blue) and retrieved pulse (red-dashed). The rightmost column displays the ideal spectrum (blue), retrieved spectrum (red-dashed), ideal spectral phase (purple), and retrieved spectral phase (yellow-dashed). Here, ideal refers to the initial pulse or spectrum before the fluctuations were applied.

Shot-to-shot energy fluctuations are extremely common in all laser systems, especially lower repetition rate systems. Figures \ref{fig:NoisyTraces}(a) through (d) show the results of a simulation where each column of the trace was generated from a pulse with identical phase and spectral content, but whose overall energy was randomly sampled from a normal distribution with a standard deviation of 25\%. The random energy fluctuations can easily be seen in the simulated trace, but they are noticeably absent in the retrieved trace. The retrieval algorithm behaves extremely favorably under such conditions as the retrieved pulse, spectrum, and spectral phase are essentially identical to the ideal pulse, spectrum and spectral phase.

Spectral phase fluctuations are also somewhat common in modern laser systems for various reasons. They can arise from pointing instabilities in the laser, or running the laser in unstable operating regimes. Figures \ref{fig:NoisyTraces}(e) through (h) show a simulation where the spectral amplitude of the pulse was identical for each column of the trace but the spectral phase of each column was the sum of the the simulated phase, $\phi_S$, and a random phase $\phi_R$ which is given by:

\begin{equation}
    \label{eq:RandPhase}
    \phi_R(\omega) = R_2 \phi_2(\omega-\omega_0)^2 + R_3 \phi_3(\omega-\omega_0)^3+R_4\phi_4(\omega-\omega_0)^4 + \sin(R_S(\omega-\omega_0))
\end{equation}

Here, $\phi_2 = 100fs^2$, $\phi_3 = 1000fs^3$, $\phi_4 = 10000fs^4$, $R_2$ through $R_4$ represent a number randomly sampled from a normal distribution with a mean of 0 and standard deviation of 1, and $R_S$ represents a number randomly sampled from a normal distribution with a mean of 0 and standard deviation of 10. Each $R_x$ was randomly sampled for each column of the trace. We find that the algorithm again behaves favorably under such strong fluctuations. Minimal deviations in the reconstructed spectrum and spectral phase are seen only near the very edges of the bandwidth. The pulse structure is accurately reconstructed with only minimal errors in the width of the side lobes.

One extremely common source of error in laser systems is pointing instabilities within the laser itself. This can slightly alter the spectral phase content of each pulse in the train due to the different paths each pulse in the train will take through the laser system. This is especially problematic with regards to grating d-scan as each pulse can enter the pulse compressor at a different incident angle, which would lead to a different grating phase being applied to each pulse. Figures \ref{fig:NoisyTraces}(i) through (l) show a simulated retrieval where the incident angle of the beam onto the compressor gratings for each column of the trace was randomly sampled from a normal distribution centered around $\theta_i = 34.05^\circ$, with a standard deviation of half a degree. The retrieval algorithm assumed a Littrow incident angle onto the gratings for the entire trace. In this case, the simulated trace does not seem to exhibit as many visible fluctuations as in the previous cases and the retrieved trace matches it extremely well. Similarly to the previous example, the retrieved spectrum and spectral phase deviate from the simulated spectrum and spectral phase very slightly near the edges of the bandwidth. Additionally, slight errors in the retrieved pulse structure can be seen in the side lobes of the temporal pulse.

Lastly, we will discuss the effects of random gain pulling on the d-scan trace. Such a phenomena is common in chirped pulse amplification systems where the amplifier is operated in an unstable regime and the long wavelength side of the chirped seed pulse experiences a disproportionate amount of gain relative to the short wavelength side randomly from pulse to pulse. This effectively randomly pulls the spectrum of the amplified pulse towards the red side of the spectrum. To simulate gain pulling, the spectral field for each column of the trace was multiplied by a weighting function to pull the spectrum towards longer wavelengths:

\begin{equation}
    \label{eq:gainPulling}
    \tilde{E}_k(\omega) = \tilde{E}_0(\omega)(1+\erf(-R_k(\omega-\omega_0)))
\end{equation}

Here, $\tilde{E}_k(\omega)$ is the spectral field of the $k^{th}$ column, $\tilde{E}_0(\omega)$ is the unperturbed spectral field, $R_k$ is a number randomly sampled from a normal distribution with a mean of 1 fs and a standard deviation of 5 fs. Figures \ref{fig:NoisyTraces}m) through p) show the results of a simulated pulse retrieval on a trace that exhibits random gain pulling in each column of the trace. The two primary effects this has are 1) the retrieved spectrum is essentially the average of all the spectra used to generate the trace, and 2) the temporal pulse is the average of all the pulses in the trace.

In effect, we see that the retrieval algorithm tends to retrieve essentially the average pulse in the pulse train when many unique pulses are present. The cases presented above are exaggerated examples of real world phenomena that are present in most modern laser systems. In reality, combinations of all these effects are present in real-world data, but as can be seen from the above examples, the algorithm performs well under such conditions and does not exhibit a misleading coherent artifact similar to how other d-scan algorithms perform under such scenarios \cite{Escoto2019-2}. While the results shown here are for the single mode algorithm, the multi-mode algorithm performs similarly under such conditions.

\bibliographystyle{unsrt}